\documentclass[usenatbib]{basi}
\usepackage{amsmath,amssymb,graphics,graphicx,verbatim,rotating,lscape,enumerate}
\begin{document}
  \title[]{Models of force-free spheres and applications \\
  to solar active regions}
 
  \author[]
         {A.~Prasad\thanks{email: \texttt{avijeet@iiap.res.in}} and
         A.~Mangalam\thanks{email: \texttt{mangalam@iiap.res.in}}\\
         Indian Institute of Astrophysics\\
        Sarjapura road, Koramangala 2nd block, Bangalore 560034, India}
 
%
 \maketitle
%
 \begin{abstract}
Here we present a systematic study of force-free field equation for simple
axisymmetric configurations in spherical geometry. 
The condition of separability of solutions in radial and angular variables leads to two classes of solutions:
linear and non-linear force-free fields. We have studied these linear solutions \citep{chandra56}
and extended the non-linear solutions given in \citet{low90} to the irreducible rational
form $n= p/q$, which is allowed for all cases of odd $p$ and to cases of $q>p$ for even $p$. We have further 
calculated their energies and relative helicities for magnetic field configurations in finite 
and infinite shell geometries. We demonstrate here a method here to be used to fit observed magnetograms as 
well as to provide good exact input fields for testing other numerical codes
used in reconstruction on the non-linear force-free fields.
\end{abstract}

\section{Introduction}\label{intro}
In systems dominated by magnetic fields in the presence of kinematic 
viscosity, linear force-free fields are the natural end states.
More general force-free fields can be obtained when the constraints of total mass,
angular momentum and helicity are put in the equations, (e.g. Finn \& Antonsen, 1983; Mangalam \&
Krishan, 2000).
There have been several attempts to 
numerically construct the full three dimensional models of coronal fields from two dimensional three 
component data available from the vector magnetograms. Some of the popular techniques include
Optimization (Wheatland et al.,2000; Wiegelmann, 2004)
Magnetofrictional (Yang et. al., 1986; McClymont et. al., 1997),
Grad-Rubin based (Amari et al., 2006; Wheatland $\&$ Leka, 2010), and
Green's function-based methods (Yan $\&$ Sakurai, 2000).
We construct analytical models of linear and non-linear axisymmetric force-free fields by solving the 
governing equations. We then take a cross-section of these 3D fields at different orientations 
to construct a library of template magnetograms corresponding to the different modes of our solutions
which can be then compared with the observed magnetograms to pick out the best fit. We apply the techniques
outlined here to magnetograms and reconstruct the coronal fields in Prasad, Mangalam \& Ravindra 
(2014, in press, henceforth referred to as PMR14), which
also contains details of the formulation presented below.

\section{Axisymmetric separable linear \& non-linear force-free fields}
\label{modes}
 The force-free magnetic field $\mathbf{B}$ is described by the equation
$
 \nabla\times\mathbf{B}=\alpha\mathbf{B}
$
An axisymmetric magnetic field can be expressed 
in terms of two scalar functions $\psi$ and $Q$ in spherical polar coordinates :
\begin{equation}
 \mathbf{B}= \frac{1}{r \sin\theta}\left(\frac{1}{r}\frac{\partial \psi}{\partial\theta}\hat{\mathbf{r}}-
\frac{\partial \psi}{\partial r}\hat{\boldsymbol{\theta}}+Q\hat{\boldsymbol{\phi}}\right).\label{spb}
\end{equation}
We try separable solutions of the form
$
 \psi=f(r)P(\mu),\quad Q=a \psi^\beta.
$
which yields
\begin{equation}
 r^2\frac{f^{\prime\prime}}{f}+(1-\mu^2)\frac{P^{\prime\prime}}{P}+a^2\beta r^2 f^{2\beta-2}P^{2\beta-2}=0.\label{fpsep}
\end{equation} 
There are two possibilities for getting separable solutions.
The third term can be a function of
\begin{enumerate}[(i)]
 \item $r$ alone, which is satisfied if $\beta=1$; these solutions were presented in Chandrasekhar (1956) and which we refer to as C modes or
\item $\mu$ alone, which is satisfied if $r^2f^{2\beta-2}=1$; these solutions were partially
explored by Low \& Lou (1990) and termed here as LL modes.
\end{enumerate}

Free energy and relative helicity are very helpful quantities for studying the
dynamics of the magnetic field configurations near active regions in the Sun.
The free energy of the system is the difference between the energies of a force-free 
field  and a potential field  in a volume. The potential field is constructed using the 
normal components of the force-free field at the boundary. The expression for free energy
$E_{free}$ is given by
$
 E_{free}=E_{ff}-E_P,
$
\begin{figure}[h!]
\centerline{\includegraphics[scale=0.4]{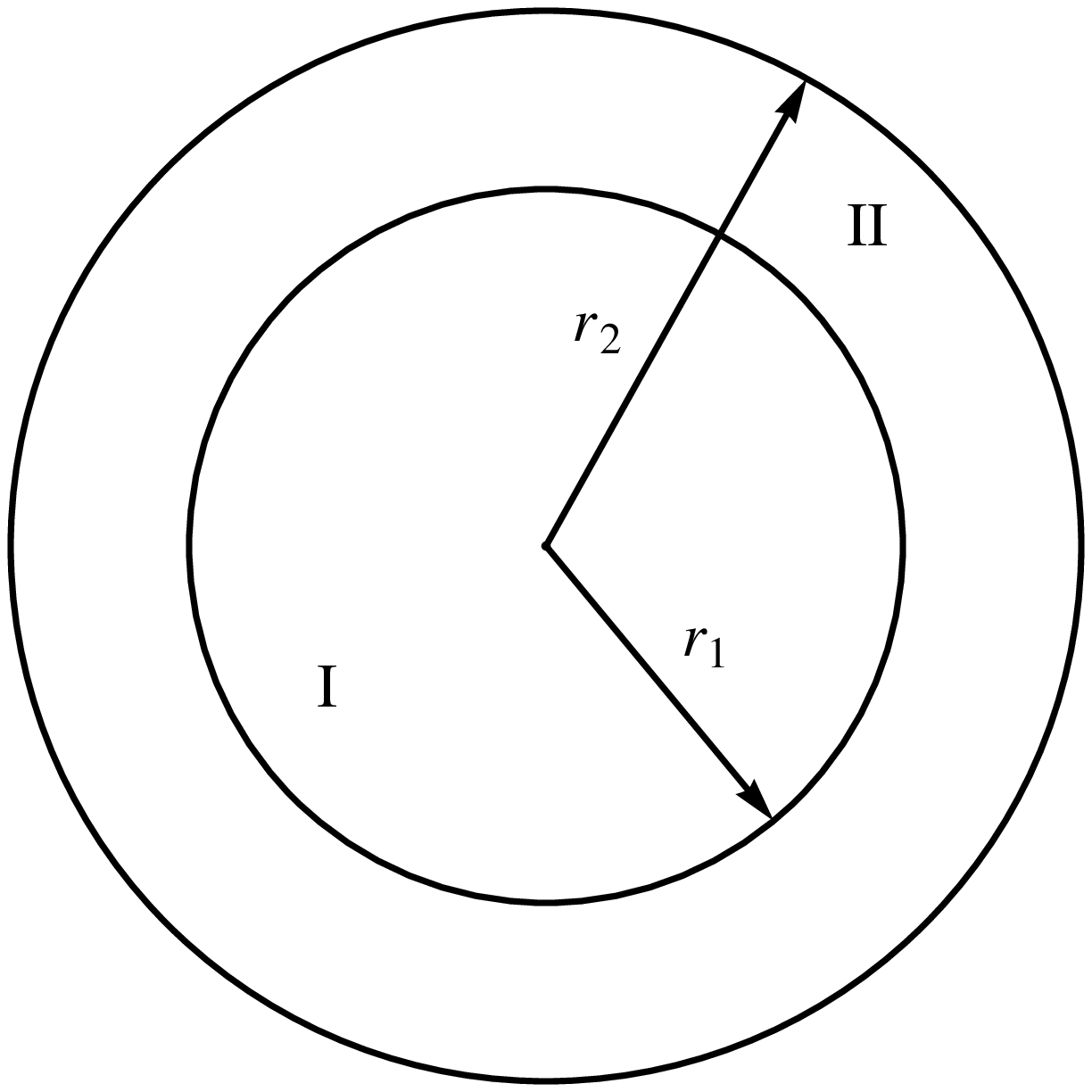} \includegraphics[scale=.25]{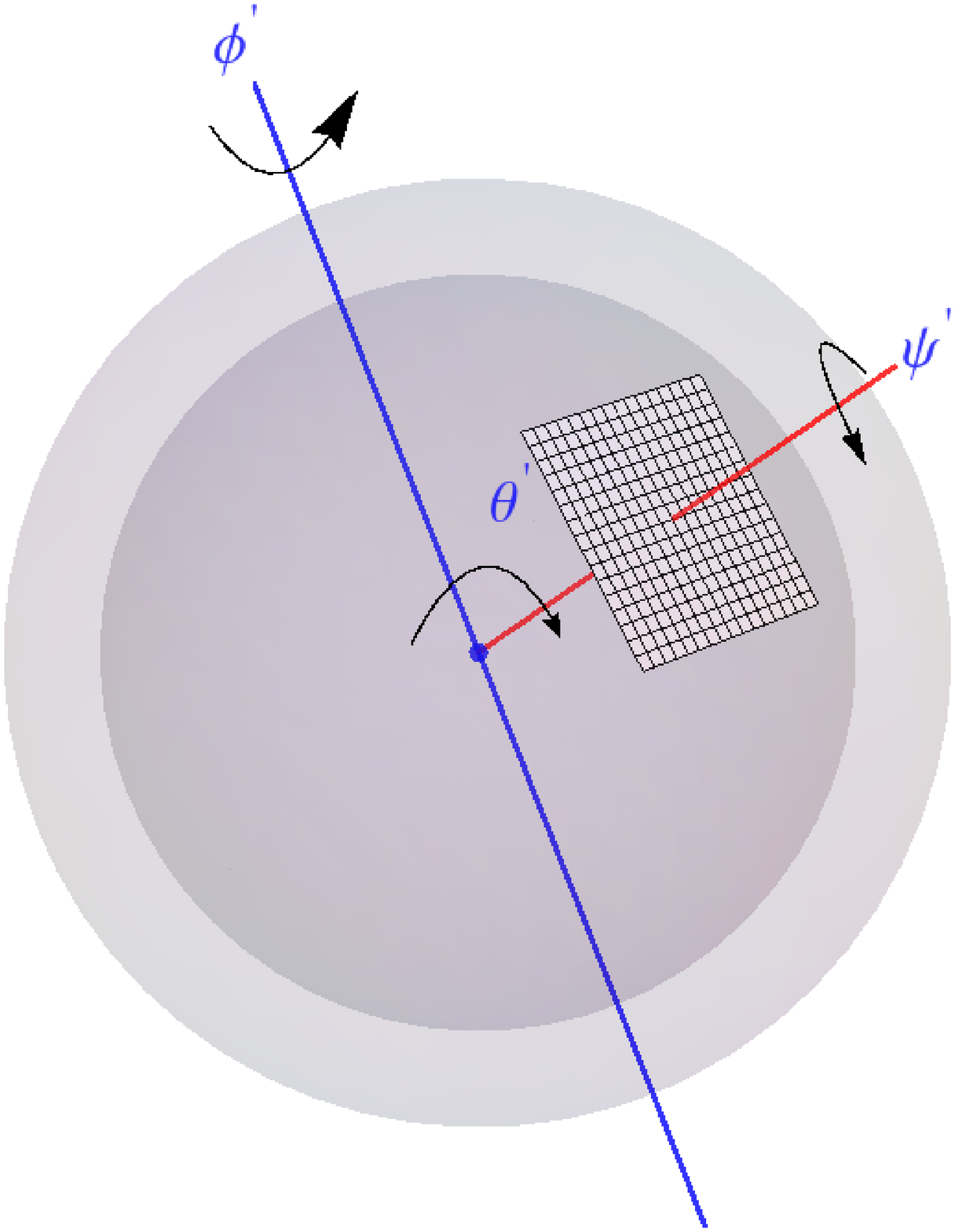}}
\caption{The left figure shows the geometry of the problem. The force-free field is first computed in the
entire region (I and II) and then corresponding potential field is 
constructed in the spherical shell between radii $r_1$ and $r_2$ 
(region II) using the normal components of the force-free field at the lower boundary, $r_1$.
The right figure shows a magnetogram which is simulated by taking a cross-section of the axisymmetric 3D force-free field
at a radius $r_1$.The magnetogram is then rotated through the Euler angles $\theta^\prime$ and $\psi^\prime$
to match the components of the observed magnetogram. The rotation $\phi^\prime$ is redundant
as the field is axisymmetric.}
\label{bound}
\end{figure}
We model the force-free field using both the linear solutions (C modes)
 and the non-linear solutions (LL modes). 
where $E_{ff}$ and $E_P$ are the energies of the force-free field and the potential field
respectively. Since the potential field is the minimum energy configuration for a given boundary 
condition, $E_{free}$ is always positive. Here we are modeling the entire active region as
as a part of a force-free sphere with an inner radius $r_1$; where the magnetogram 
measurements are available and an outer radius $r_2$ as shown in the left panel of Fig.\ref{bound}.
 
\textbf{C modes}: with the condition $\beta=1$, 
we get 
 $
\displaystyle{
r^2\frac{f^{\prime\prime}}{f}+a^2 r^2=n(n+1)
}
$
as the equation for the radial part where $n$ is a constant,
the solution to which is given by
\begin{equation}
 f_n(r)=c_1\sqrt{r}J\left[(1+2n)/2,ar\right]+c_2\sqrt{r}Y\left[(1+2n)/2,ar\right].
\end{equation}
The angular part is given by the following equation
$
 (1-\mu^2)\frac{P^{\prime\prime}}{P} =-n(n+1)
$
whose solution is given by
$ 
P(\mu)= (1-\mu^2)^{1/2}P_n^1(\mu).
$
\textbf{LL modes}: the second condition $r^2f^{2\beta-2}=1$ implies $\beta=(n+1)/n$ for the functional form
$
f(r)=r^{-n},
$
The differential equation for the angular part then becomes
\begin{equation}
(1-\mu^2)\frac{P^{\prime\prime}}{P}+a^2\frac{n+1}{n}P^{1+2/n}+n(n+1)P=0.
\label{peq}
\end{equation}
The above equation is solved numerically  as no general closed form 
is known for all values of $n$. For a given value of $n$, we get different modes
for different eigenvalues of $a_{n,m}$ satisfying the above equation for a given boundary condition.
Low $\&$ Lou(1990) were able to solve the above equation only for $n=1$ due to singular 
nature of the solutions. We were able to tackle this problem for higher values of $n$ through the transformation
$
 P(\mu)= (1- \mu^2)^{1/2} F(\mu),
$
by which eqn (\ref{peq}) now stand as
\begin{equation}
 (1-\mu^2)F^{\prime\prime}(\mu)-2\mu F^\prime(\mu)+\left[n(n+1)-\frac{1}{(1-\mu^2)}\right]
F(\mu)+a^2\frac{(n+1)}{n}F^{\frac{(n+2)}{n}}(1-\mu^2)^\frac{1}{n}=0.
\label{feq}
\end{equation}
We were able to obtain an infinite set of solutions for all rational values of $n=\displaystyle{\frac{p}{q}}$.
For even values $p$, solutions exist if $F(\mu)>0$ in the domain $-1 \leq \mu \leq 1$.
In this case we get
$
\displaystyle{
 \alpha=\frac{a (1+1/n)(1-\mu^2)^{1/2n}F^{1/n}}{r^n}.
}
$
The physical quantities of interest such as the force-free energy, free energy, $E_{ff}$, $E_{free}$
and the relative helicity $H_{rel}$ are calculated for the region above the magnetogram.
The formulary of the results for 
the C and LL modes are presented in Table \ref{t:formulae}. 
The details of derivation are presented in PMR14.
\begin{table}
\begin{tabular}{|l|}
\hline
\centerline{C modes}\\ \hline
$
\mathbf{B}(r<r_2)=\left(-\frac{1}{r^2}\frac{\partial}{\partial\mu}[S_mr^2(1-\mu^2)],
\frac{-1}{r\sqrt{(1-\mu^2)}}\frac{\partial}{\partial r}[S_mr^2(1-\mu^2)],
\alpha r\sqrt{(1-\mu^2)}S_m\right);
\mathbf{A}(r<r_2)=\mathbf{B}/\alpha
$

\\
\\
$
\chi_{m+1}(r_1)=\frac{(m+1)(m+2)}{r_1^{3/2}}J_{m+3/2}(\alpha r_1);
a_{m+1}=\frac{\chi_{m+1}(r_1)}{(m+1)}\frac{r_1^{m+3}}{ r_1^{2m+3}-r_2^{2m+3}};
b_{m+1}=\frac{(m+1)}{(m+2)}a_{m+1} r_2^{(2m+3)}
$

\\\\ 
$
\mathbf{B}_P(r_1<r<r_2)=\left(\left[(m+1) a_{m+1} r^{m}-\frac{(m+2)b_{m+1}}{r^{m+3}}\right]P_{m+1}(\mu),\right.
\left.-(1-\mu^2)^{1/2}\left[ a_{m+1} r^{m}+\frac{b_{m+1}}{r^{m+3}}\right]\frac{\partial P_{m+1}}{\partial \mu},0\right)
$

\\ \\
$
\mathbf{A}_P(r_1<r<r_2)=\left(0,0,(1-\mu^2)^{1/2} P^\prime_l(\mu)\left[\frac{ a_l r^l}{l+1}-\frac{b_l}{lr^{l+1}}\right]\right)
$

\\ \\
$
E_{ff}(\alpha,n,m, r_1, r_2)=\frac{(m+1)(m+2)}{2(2m+3)}\Bigl[(m+1)(m+2)\int_{r_1}^{r_2}\frac{J^2_{m+3/2}(\alpha r)}{r}\textrm{d} r+r^{3/2}J_{m+3/2}(\alpha r)|^{r_1}_{r_2}$
\\\\
$
~~~~~~~~~~~~~~~~~~~~~~~~~~~~-\int_{r_1}^{r_2}r^{1/2}J_{m+3/2}(\alpha r)\textrm{d} r+\alpha^2\int_{r_1}^{r_2}r J^2_{m+3/2}(\alpha r)\textrm{d} r\Bigr]
$
\\ \\
$
E_{pot}(m,r_1,r_2)=\frac{1}{2(2m+3)}\int_{r_1}^{r_2}\Bigl[\left((m+1)a_{m+1}r^{m+1}-\frac{(m+2)b_{m+1}}{r^{m+2}}\right)^2$
\\\\
$
~~~~~~~~~~~~~~~~~~~+(m+1)(m+2)\left(a_{m+1}r^{m+1}+\frac{b_{m+1}}{r^{m+2}}\right)^2\Bigr]\textrm{d}r
$

\\ \\
$
H_{rel}(\alpha,n,m,r_1,r_2)=\frac{E_{ff}(\alpha,n,m)}{\alpha}+{1 \over \alpha}\Bigl[\frac{2(m+1)(m+2)}{(2m+3)}\Bigl[\left(a_{m+1} r^{m+1} + \frac{b_{m+1}}{r^{m+2}}\right)\left(r^{1/2} J_{m+3/2}(\alpha r) \right)\Bigr]_{r_1}^{r_2}
$
\\\\
$~~~~~~~~~~~~~~~~~~~~~~~~~~~
-\int_{r_1}^{r_2}r^{1/2} J_{m+3/2}(\alpha r)\left[(m+1)a_{m+1} r^m-\frac{(m+2)b_{m+1}}{r^{m+3}}\right]
$
 \\\\
\hline
\centerline{LL modes} \\ \hline
$
\mathbf{B}(r<r_2)= \left(\frac{-1}{r^{n+2}}\frac{\partial P}{\partial\mu},
\frac{n}{r^{n+2}}\frac{P}{(1-\mu^2)^{1/2}},\frac{a}{r^{n+2}}\frac{P^{(n+1)/n}}{(1-\mu^2)^{1/2}}\right);
\mathbf{A}(r<r_2)=\left(0,\frac{-a}{n r^{n+1}}\frac{P(\mu)^{(n+1)/n}}{(1-\mu^2)^{1/2}},\frac{1}{r^{n+1}}\frac{P(\mu)}{(1-\mu^2)^{1/2}}\right)
$
\\\\
$
a_l=0,\quad b_l=\frac{2l+1}{2(l+1)}r_1^{l-n}\int_{-1}^1\frac{\partial P}{\partial \mu}P_l(\mu)d\mu
$
 \\ \\
$
\mathbf{B}_P(r_1<r<r_2)=\left(\sum_{l=0}^{\infty}-(l+1)\frac{b_l}{r^{l+2}}P_l(\mu),\sum_{l=0}^{\infty}\frac{-b_l}{r^{l+2}}
(1-\mu^2)^{1/2}\frac{\partial P_l}{\partial \mu},0\right).
$
 \\ \\
$
\mathbf{A}_P(r_1<r<r_2)=\left(0,0,(1-\mu^2)^{1/2} P^\prime_l(\mu)\left[\frac{ a_l r^l}{l+1}-\frac{b_l}{lr^{l+1}}\right]\right)
$
 \\ \\
$
E_{ff}(n,m, r_1)=\frac{1}{4(2n+1)r_1^{2n+1}}\int_{-1}^1\mathrm{d}\mu
\left[ P^\prime(\mu)^2+\frac{n^2 P(\mu)^2}{1-\mu^2}+\frac{a^2 P(\mu)^{(2n+2)/n}}{1-\mu^2}\right]
$
\\ \\
$
E_{pot}(l,r_1)=\sum_{l=0}^\infty\frac{b_l^2(l+1)^2}{2(2l+1)^2r_1^{2l+1}};
H_{rel}(n,m,r_1)=-2\pi a \sum_{l=0}^\infty \int_{-1}^1  \frac{b_l}{n l r_1^{n+l}}P^{1+1/n} \frac{\textrm{d} P_l}
{\textrm{d} \mu}\textrm{d}\mu
$
 \\\\
\hline
\end{tabular}
\caption{Formulary for the various quantities calculated for the C and LL modes. 
$\mathbf{B}$ and $\mathbf{A}$ denote the force-free magnetic field and its corresponding 
vector potential. The same quantities for the potential field are denoted by $\mathbf{B}_P$
and $\mathbf{A}_P$ respectively. $E_{ff}$, $E_{pot}$, $E_{free}$ and $H_{rel}$
are the force-free energy, potential energy, free energy and the relative helicity 
of the magnetic field configuration respectively. See PMR14 for details of the derivation.}
\label{t:formulae}
\end{table}
\section{Simulation of magnetograms}
\label{s:simulation}
\begin{figure}[h!]
\centerline{\includegraphics[scale=.4]{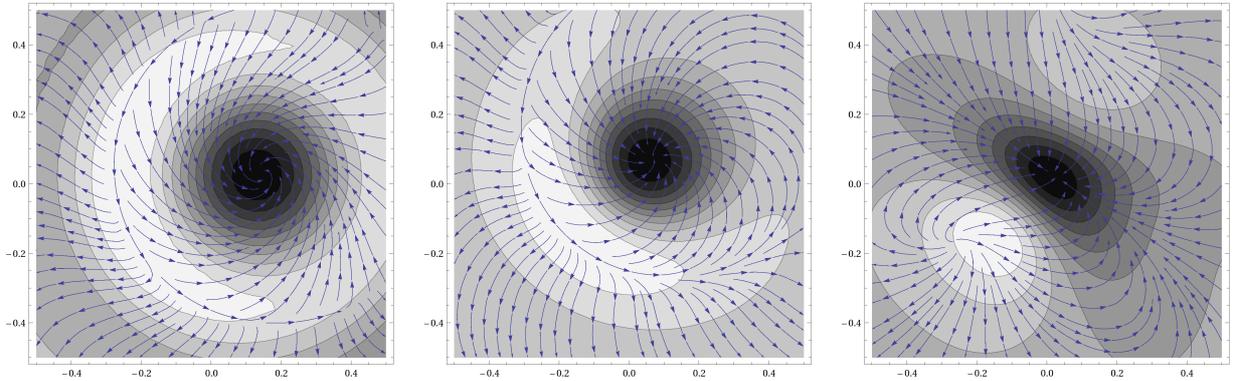}}
\caption{Examples of magnetogram sections are presented in the figure for C (left) and LL  
(center and right) modes. The parameters for the C modes are ($\alpha,n,m,r_1,r_2,\theta,\phi$)=
(-10.95,2,10,1.35,1.85,0.3,-0.12) and the parameters for the LL modes are given by
($n,m,r_1,\theta,\phi$)=(3,2,0.5,0.31,4.1) and (1.4,2,0.3,1.75,4.14) for the center and right panel
respectively. The left and the center correspond to a double polarity region whereas the right panel
represents a triple polarity region.}
\label{f:template}
  \end{figure}

The following steps are involved in the simulation of the magnetogram:
(i)  An operator $\Lambda(\theta', \phi')$ is used for the Euler rotation to find the orientation
 of the magnetogram, see Fig. (\ref{bound}, right panel). The expression for $\Lambda$ is given by
\begin{equation}
 \Lambda(\theta^\prime,\psi^\prime)=\begin{bmatrix}\cos\psi^\prime  &\cos\theta^\prime\sin\psi^\prime  &\sin\psi^\prime\sin\theta^\prime  \\
 -\sin\psi^\prime  &\cos\theta^\prime\cos\psi^\prime   &\cos\psi^\prime\sin\theta^\prime \\ 0 &-\sin\theta^\prime   & \cos\theta^\prime \end{bmatrix}.
\end{equation}
(ii) An operator $S$ is used for transformation of the coordinates from cartesian $(x,y,z)$ 
to spherical $(r,\theta,\phi)$.
(iii) An operator $T(\theta,\phi)$ is used to transform the magnetic field vector $\mathbf{B}$ from
spherical to cartesian coordinates.
\begin{equation}
 T=\begin{bmatrix} \sin\theta\cos\phi & \cos\theta\cos\phi & -\sin\phi \\ \sin\theta\sin\phi  & \cos\theta\sin\phi  &\cos\phi \\
\cos\theta & -\sin\theta & 0 \end{bmatrix}
\end{equation}
(iv) A cartesian point on the magnetogram $\mathbf{x}_C \equiv (x,y,z)$ is first rotated through the
inverse of $\Lambda$ and then converted to spherical coordinates
 $  \mathbf{x}_S \equiv (r,\theta,\phi)$ through the operation of $S$
such that
$
\mathbf{x}_S=  S \left( \Lambda^{-1}(\theta', \psi')\mathbf{x}_C \right).
$
 (v) We then evaluate the magnetic field in spherical coordinates with
$\mathbf{B}_S(\mathbf{x}_S)$ and then convert the components of magnetic 
field to cartesian through the $T$ and obtain the correct orientation by 
the operator $\Lambda$ given by
$
\mathbf{B}_C [\mathbf{x}_C]= \Lambda(\theta', \psi') T \left( \mathbf{B}_S \left[ \mathbf{x}_S \right] \right).
$
In Fig. \ref{f:template}, we show realizations of magnetograms thus constructed for the 
cases of C and LL modes. These templates can then be compared with the available 
photospheric vector magnetograms and thus providing a full 3 dimensional and 3 component information
of the coronal magnetic fields. Such studies using photospheric magnetograms obtained from HINODE
are presented in PMR14. 
\section{Summary \& Conclusions}
We have shown that there are two solutions possible (albeit known already and denoted here as C and LL)
from the separability assumption. We calculate the energies and relative helicity of the allowed force free
fields in a shell geometry. For the LL mode we were able to extend the solution set obtained in \citet{low90} from $n=1$ to  
all rational values of $n=\displaystyle{\frac{p}{q}}$ by solving the eqn (\ref{feq}) for all 
cases of odd $p$ and for cases of $q>p$ for even $p$, in effect extending solution to practically all $n$. 
The results are presented in Table \ref{t:formulae}. 
The LL solution of ($n=1$ in \citet{low90} and $n=5,7,9$ (odd cases) in \citet{flyer04})
have been extended here to the cases of nearly all $n$.  
The topological properties of these extended solutions can be further studied by considering other
boundary conditions.The analytic solutions for LL suffer from the problem of a singularity at 
the origin which render them unphysical; this implies that more realistic boundary
conditions are necessary. To learn more about the evolution and genesis of these structures, it would be useful to carry
out dynamical simulations allowing for foot point motions with the analytic input fields constructed above 
to study how the non-linearity develops; a stability analysis of the non-linear modes would also useful tool 
(\citet{berger85} has analyzed the linear constant $\alpha$ case). Clearly, these are difficult mathematical 
problems to be addressed in the future. We explore fits of these solutions to HINODE magnetograms of 
NOAA AR 10930, 10923 and 10933 and obtain the best fits to C and LL modes using the procedure discussed above in PMR14.

We would also like to thank Y. Yan and M. Wheatland for helpful discussions. 
AP would like to thank CSIR for the SPM fellowship.

\end{document}